\documentclass[journal]{IEEEtran}

\usepackage{amsmath}
\usepackage{amssymb}
\usepackage{bm} % 处理加粗数学符号

\usepackage{graphicx}

\usepackage{float}

\usepackage{xcolor,stfloats}
\usepackage{lipsum}

\usepackage{algorithm, algpseudocode}
\usepackage{ifpdf}
\usepackage{cite}
\usepackage{hyperref}
\ifCLASSINFOpdf
\else
\fi
\usepackage{amsmath}
\usepackage{url}
\begin{document}

% ————————————————————————————————————————————————————
% ————————————————————————————————————————————————————
% ———————————————————————————————————————————————————— title
% ———————————————————————————————————————————————————— paper winwinwin!
% ———————————————————————————————————————————————————— ACCEPT!!!
% ————————————————————————————————————————————————————

\title{Semantic Learning for Molecular Communication in Internet of Bio-Nano Things}


\author{Hanlin~Cai,~\IEEEmembership{Student~Member,~IEEE,}
        and~Ozgur~B.~Akan,~\IEEEmembership{Fellow,~IEEE}% <-this % stops a space

% \thanks{Manuscript received December 20, 2024.}
% (Corresponding authors: Ozgur~B.~Akan).}
\thanks{Hanlin Cai is with the Internet of Everything Group, Electrical Engineering Division, Department of Engineering, University of Cambridge, CB3 0FA Cambridge, U.K. (e-mail: hc663@cam.ac.uk).}% <-this % stops a space
\thanks{Ozgur B. Akan is with the Internet of Everything Group, Electrical Engineering Division, Department of Engineering, University of Cambridge, CB3 0FA Cambridge, U.K., and also with the Center for neXt-Generation Communications (CXC), Department of Electrical and Electronics Engineering, Koç University, 34450 Istanbul, Turkey (e-mail: oba21@cam.ac.uk).}% <-this % stops a space
% \thanks{Digital Object Identiﬁer}

}



\markboth{Accepted by the 9th Workshop on Molecular Communications, April~2025}%
{}



% \IEEEspecialpapernotice{(Invited Paper)}

\maketitle


\begin{abstract}

Molecular communication (MC) provides a foundational framework for information transmission in the Internet of Bio-Nano Things (IoBNT), where efficiency and reliability are crucial. However, the inherent limitations of molecular channels, such as low transmission rates, noise, and inter-symbol interference (ISI), limit their ability to support complex data transmission. This paper proposes an end-to-end semantic learning framework designed to optimize task-oriented molecular communication, with a focus on biomedical diagnostic tasks under resource-constrained conditions. The proposed framework employs a deep encoder-decoder architecture to efficiently extract, quantize, and decode semantic features, prioritizing task-relevant semantic information to enhance diagnostic classification performance. Additionally, a probabilistic channel network is introduced to approximate molecular propagation dynamics, enabling gradient-based optimization for end-to-end learning. Experimental results demonstrate that the proposed semantic framework improves diagnostic accuracy by at least 25\% compared to conventional JPEG compression with LDPC coding methods under resource-constrained communication scenarios.

\end{abstract}

% Note that keywords are not normally used for peerreview letters.

% \fi
\IEEEpeerreviewmaketitle

\vspace{-6pt}

\section{Introduction}

\IEEEPARstart{M}{olecular} communication (MC) has emerged as a promising paradigm for information exchange in environments where traditional electromagnetic (EM)-based communication systems encounter fundamental limitations. Unlike EM waves, which suffer from severe attenuation and interference in biological and fluidic environments, MC relies on the controlled release, propagation, and detection of molecules to encode and transmit information \cite{akan2016fundamentals}. This approach is particularly well-suited for applications in the Internet of Bio-Nano Things (IoBNT), where micro- and nanoscale devices operate in biological systems \cite{dhok2021cooperative}. Key IoBNT applications include disease diagnostic, targeted drug delivery, and real-time health monitoring, where MC provides a biocompatible and energy-efficient communication mechanism \cite{jamali2019channel}.

Despite its potential, the practical deployment of MC in IoBNT faces significant challenges, including low data rates, severe inter-symbol interference (ISI), and high susceptibility to noise. These impairments substantially limit the molecular channel's ability to support complex data transmission, which is critical for biomedical applications such as disease diagnosis and physiological signal monitoring \cite{xiao2023really}. Given the stochastic nature of molecular propagation, addressing these challenges requires novel approaches to enhance communication efficiency while ensuring robustness under dynamic and uncertain channel conditions \cite{baydas2023estimation}.

To overcome these limitations, incorporating semantic processing into communication systems has emerged as a promising solution for optimizing resource-constrained environments by prioritizing task-relevant information over conventional bit-level accuracy \cite{getu2025semantic}. In \cite{bourtsoulatze2019deep}, a semantic-based joint source-channel coding (JSCC) framework was introduced to directly map source data to channel symbols, eliminating the need for separate compression and error correction. By jointly optimizing encoding and decoding, JSCC demonstrated enhanced robustness against noise and bandwidth constraints in wireless communication, ensuring graceful performance degradation under varying channel conditions. The work in \cite{li2022domain} extended semantic communication to the IoBNT domain by integrating domain knowledge into the encoding process, improving efficiency in biologically constrained environments with strict energy and resource limitations. Furthermore, \cite{yukun2024building} investigated the integration of semantic communication with molecular systems. This study introduced an end-to-end training approach to enhance communication reliability under stochastic propagation effects, demonstrating the feasibility of semantic encoding in molecular channels.

Although semantic-based methods have been explored in molecular communication, existing approaches struggle to map task-relevant information into physically transmittable molecular parameters while accounting for the stochastic and non-differentiable nature of molecular propagation. Moreover, the lack of a structured mapping between high-level semantic information and molecular transmission parameters limits the adaptability and transferability of current models across dynamic channel conditions and diverse IoBNT tasks.

In this paper, we propose an end-to-end semantic molecular communication framework using a deep encoder-decoder architecture to extract, quantize, and decode task-relevant semantic features. We introduce a quantization function to optimize the semantic-to-physical mapping and enhance system transferability. To achieve channel differentiability, we further propose a probabilistic channel network that models the stochastic dynamics of molecular propagation. This integration facilitates end-to-end training and dynamic adaptation to channel conditions. Unlike conventional methods focused on bit-level transmission, our method prioritizes task-relevant semantics, demonstrating superior efficiency and robustness over traditional methods in diagnostic classification tasks.

%%%%%%%%%%%%%%%%%%%%%%%%%%%%%%%%%%%%%%%%%%%%%%%%%%%%%%%%%%%%%%%
%%%%%%%%%%%%%%%%%%%%%%%%%%%%%%%%%%%%%%%%%%%%%%%%%%%%%%%%%%%%%%%

\begin{figure} %[h]
    \centering
    \includegraphics[width=1\linewidth]{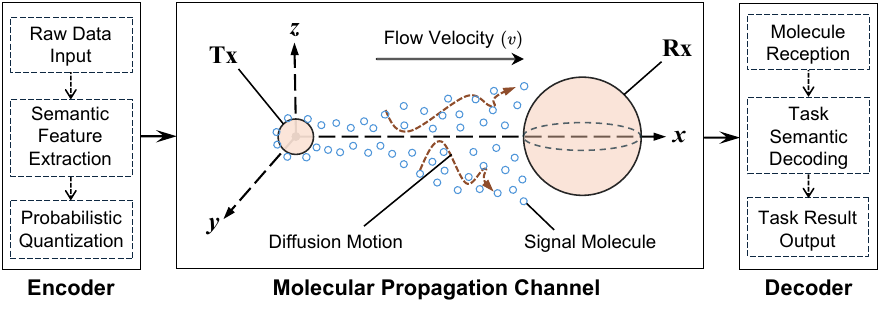}
    \caption{Illustration of the semantic molecular communication framework.}
    % \caption{Molecular propagation channel with a point Tx and a spherical Rx.}
    \label{fig:channel}
    \vspace{-10pt}
\end{figure}

\section{Semantic Molecular Communication System}
In this paper, we consider a SISO molecular communication system operating in an unbounded, three-dimensional environment with a constant uniform flow velocity, as illustrated in Fig. \ref{fig:channel}. The transmitter (Tx) and receiver (Rx) are assumed to be perfectly synchronized, ensuring precise alignment for each symbol transmission. The transmitter encodes the input data by instantaneously releasing at most \( n_m \) identical molecules at the beginning of each symbol slot with duration \( t_s \).

\subsection{Molecular Propagation Model}

The molecular communication (MC) system under consideration consists of a point Tx positioned at \(\mathbf{d}_{\mathrm{Tx}} = [0, 0, 0]\) and a spherical Rx centered at \(\mathbf{d}_{\mathrm{Rx}} = [R, 0, 0]\). Molecules propagate through the medium under the influence of diffusion and advection, where a uniform drift velocity \(\mathbf{v} = [v, 0, 0]\) acts along the \(x\)-axis. The molecular concentration \(\Phi(\mathbf{d}, t)\) at position \(\mathbf{d}\) and time \(t\) follows the advection–diffusion equation:
\begin{equation}
\frac{\partial \Phi(\mathbf{d}, t)}{\partial t} = D \nabla^2 \Phi(\mathbf{d}, t) - \nabla \bigl( \mathbf{v} \Phi(\mathbf{d}, t) \bigr),
\end{equation}
where \( D_c \) is the diffusion coefficient, \(\nabla^2\) represents the Laplace operator, and \(\nabla\) denotes the gradient operator. The first term describes molecular diffusion, while the second term accounts for the drift effect along the velocity field. Based on the analytical derivation in~\cite{jamali2019channel}, the probability of molecular capture at the Rx is given by:
\begin{equation}
P(t) = \frac{V_r}{\bigl(4\pi D_c\,t\bigr)^{3/2}} 
\exp\!\Bigl(-\frac{\bigl(R - v\,t\bigr)^2}{4\,D_c\,t}\Bigr),
\end{equation}
where \( V_r = \tfrac{4\pi r^3}{3} \) represents the Rx volume, \( R \) denotes the distance between the Tx and Rx, and \( t \) is the elapsed time. Since each signal molecule is transmitted independently, the number of molecules detected by Rx follows a binomial distribution. However, When \( n_m \) is sufficiently large, the distribution can be further approximated by a Gaussian distribution as \cite{yukun2024building}:
\begin{equation}
N(n_m, t) \sim B\big(n_m, P(t)\big) \sim \mathcal{N}\Big(n_m P(t), n_m P(t)\big(1 - P(t)\big)\Big),
\end{equation}
where \( B(\cdot) \) and \( \mathcal{N}(\cdot) \) denote the binomial and Gaussian distributions, respectively. During the communication process, molecules released in prior time slots may arrive at the receiver due to the uncertainty introduced by molecular diffusion. This effect, known as inter-symbol interference (ISI), is typically negligible when the drift velocity significantly dominates the Brownian diffusion. However, when diffusion becomes the predominant propagation mode, ISI can degrade the system's communication performance significantly. Additionally, the MC channel may introduce Gaussian noise due to molecular decomposition or emissions from other nano-machines \cite{zhu2023evolutionary}. This noise is modeled as \( N_{\text{noise}} \sim \mathcal{N}(0, \sigma_n^2) \). Considering both ISI and noise, the total number of molecules observed by the Rx at the \( j \)-th time slot can be expressed as:
\begin{equation}
N = W[j]N(n_m, t) + \sum_{i=1}^{\lambda} W[j-i] N(n_m, t + it_s) + N_{\text{noise}},
\end{equation}
where \( W[j] \) represents the transmitted symbol bit at the \( j \)-th time slot, and \( \lambda \) denotes the channel memory length, capturing the residual influence of prior transmissions. For analytical tractability, we assume \( \lambda = 1 \), considering only ISI contributions from molecules released in the immediately preceding time slot. Based on this, the signal-to-interference ratio (SIR) of the proposed model can be expressed as:
\begin{equation}
SIR = \frac{W[j] N(n_m, t)}{\sum_{i=1}^{\lambda} W_{(j - i)} N(n_m, t + it_s) + N_{\text{noise}}}.
\end{equation}

\subsection{Semantic Encoder and Decoder}
The semantic coding design of the molecular communication system is structured to accommodate a wide range of input data types, provided they align with the semantic objectives of the system. The input data, denoted as \( \chi \), can represent diverse forms of information, such as biomedical images or environmental sensory data, depending on the specific application context. In this work, \( \chi \in \mathbb{R}^{H \times W \times C} \) represents biomedical images used for gastrointestinal disease diagnostic tasks in the IoBNT. The proposed framework integrates semantic feature extraction, quantization, and decoding to enable robust end-to-end learning in molecular communication systems. This section provides a detailed explanation of the proposed system, explaining the functions of the encoder and decoder.

\subsubsection{Semantic Feature Extraction}
The encoder transforms the input biomedical image \( \chi \) into a lower-dimensional semantic representation \( \mathcal{F} \in \mathbb{R}^k \) through the mapping \( \mathcal{F} = f_{\theta}(\chi) \). Here, \( f_{\theta}(\cdot) \) represents a convolutional neural network (CNN) augmented with a final linear transformation layer. This design extracts task-relevant semantic features while minimizing redundancy, providing a continuous, unnormalized representation \( \mathcal{F} \) suitable for subsequent processing. The semantic features \( \mathcal{F} \) encapsulate high-level abstractions of the input image \( \chi \), such as diagnostically significant patterns or regions indicative of disease severity. The encoder consists of five convolutional layers, each followed by batch normalization to stabilize training and \textit{LeakyReLU} activation functions to introduce non-linearity. This hierarchical design progressively reduces the spatial dimensions of the input image while increasing the abstraction level of the extracted features.

\subsubsection{Probabilistic Quantization}
To transform the semantic features \( \mathcal{F} \) into a normalized vector of channel input symbols \( W = [W_1, W_2, \dots, W_k] \), the quantization function \( W = Q_{\beta}(\mathcal{F}) \) is employed as a critical intermediate step. Here, \( Q_{\beta}(\cdot) \) is implemented as a three-layer fully connected neural network that maps each component \( \mathcal{F}_i \) of the semantic feature vector to a corresponding element \( W_i \in [0, 1] \) in the normalized output vector \( W \). In this framework, the actual number of molecules released by the transmitter at each symbol slot is given by \( (W_i \times n_m )\), where \( n_m \) denotes the maximum molecular release capacity in each symbol duration. This formulation guarantees that the molecular transmission is dynamically modulated based on the encoded semantic information while adhering to physical constraints on molecular release.

\subsubsection{Task-Specific Decoding}
The decoder directly maps the received channel output symbols \( W_{\text{Rx}} \) to the task-specific output \( y \), where \( y \) represents the predicted probability distribution over classes in the diagnostic classification task. The decoding process is formulated as: \( y = g_{\psi}(W_{\text{Rx}}) \), where \( g_{\psi}(\cdot) \) is designed as a fully connected neural network with three layers to process the normalized molecular transmission parameters \( W_{\text{Rx}} \) into the semantic output \( y \). The final layer employs a \textit{Softmax} activation function to produce a probability distribution over the task-specific output space. To optimize the framework for the diagnostic classification, the cross-entropy loss function is employed, which is minimized during training: 
\begin{equation}
\label{equ_l_ce}
\mathcal{L}_{\mathrm{ED}}=-\sum_{i=1}^k z_i \log \left(y_i\right),~~
(\theta^*, \beta^*, \psi^*) = \arg \min_{\theta, \beta, \psi}\mathcal{L}_{\mathrm{ED}},
\end{equation}
where \( z_i \) and \( y_i \) represent the one-hot encoded ground truth and predicted probability distributions, respectively. This end-to-end optimization enables joint training of the encoder and decoder, maximizing task accuracy while enhancing robustness against molecular channel impairments.

%%%%%%%%%%%%%%%%%%%%%%%%%%%%%%%%%%%%%%%%%%%%%%%%%%%%%%%%%%%%%%%
%%%%%%%%%%%%%%%%%%%%%%%%%%%%%%%%%%%%%%%%%%%%%%%%%%%%%%%%%%%%%%%

\subsection{Channel Network}
The channel network is a vital component of the communication framework, serving as a probabilistic model to capture the stochastic behavior of molecular communication channels. These channels are inherently random due to phenomena such as noise, molecular diffusion, and ISI, all of which can significantly distort the transmitted channel symbols \( W \). By modeling the conditional probability distribution of the received symbols \( W_{\text{Rx}} \), the channel network effectively addresses these challenges and enables robust end-to-end optimization. To capture the stochastic transformations introduced by the channel, the channel network models the conditional distribution of \( W_{\text{Rx}} \) as a mixture of Gaussian distributions:
\begin{equation}
p(W_{\text{Rx}}|W) = \sum_{i=1}^{h} \pi_i(W) \varphi_i(W_{\text{Rx}}|W),
\end{equation}
where \( h=2 \) represents the number of Gaussian components, \( \pi_i \) are the mixing coefficients that satisfy \( \sum_{i=1}^{h} \pi_i(W) = 1 \), and \( \varphi_i(W_{\text{Rx}}) \) denotes the \( i \)-th Gaussian kernel:
\begin{equation}
\varphi_i(W_\text{Rx}|W) = \frac{1}{\sqrt{2\pi~\sigma^2_i(W)}} \exp \frac{-\left\|(W_{\text{Rx}} - \mu_i(W)\right\|^2}{2\sigma^2_i(W)},
\end{equation}
where \( \mu_i \) and \( \sigma_i^2 \) representing the mean and variance of the \( i \)-th Gaussian component, respectively. The channel network is implemented with five fully connected layers designed to estimate the parameters of the Gaussian mixture model, including the means (\( \mu_i \)), variances (\( \sigma_i^2 \)), and mixing coefficients (\( \pi_i \)). These learned parameters define the conditional distribution \( p(W_{\text{Rx}}|W) \), facilitating accurate modeling of the stochastic channel effects and enabling robust decoding of the transmitted symbols. The channel network is trained separately using randomly generated channel symbols vectors \( W \) and their corresponding received vectors \( W_{\text{Rx}} \). The channel network learns the conditional probability \( p(W_{\text{Rx}}|W) \), modeled as a Gaussian mixture distribution, by minimizing the negative log-likelihood loss \cite{garcia2022model}:
\begin{equation}
\label{equ_l_cn}
\mathcal{L}_{\text{CN}} = -\frac{1}{k} \sum_{j=1}^k \log\left( \sum_{i=1}^h \pi_i(W[j]) \varphi_i(W_{\text{Rx}}[j]|W[j]) \right),
\end{equation}
where \( W[j] \) and \( W_{\text{Rx}}[j] \) denote the transmitted and received channel symbols at the \( j \)-th instance. Upon completion of training, the channel network’s parameters are fixed, and the network is incorporated as a static component during the subsequent joint optimization of the encoder and decoder. This ensures that the encoder and decoder can adapt their parameters to optimize task performance based on the fixed approximation of the molecular channel. The training process is considered complete when (\ref{equ_l_ce}) and (\ref{equ_l_cn}) converge.

%%%%%%%%%%%%%%%%%%%%%%%%%%%%%%%%%%%%%%%%%%%%%%%%%%%%%%%%%%%%%%%
%%%%%%%%%%%%%%%%%%%%%%%%%%%%%%%%%%%%%%%%%%%%%%%%%%%%%%%%%%%%%%%

\begin{figure}[t]
    \centering
    \includegraphics[width=0.832\linewidth]{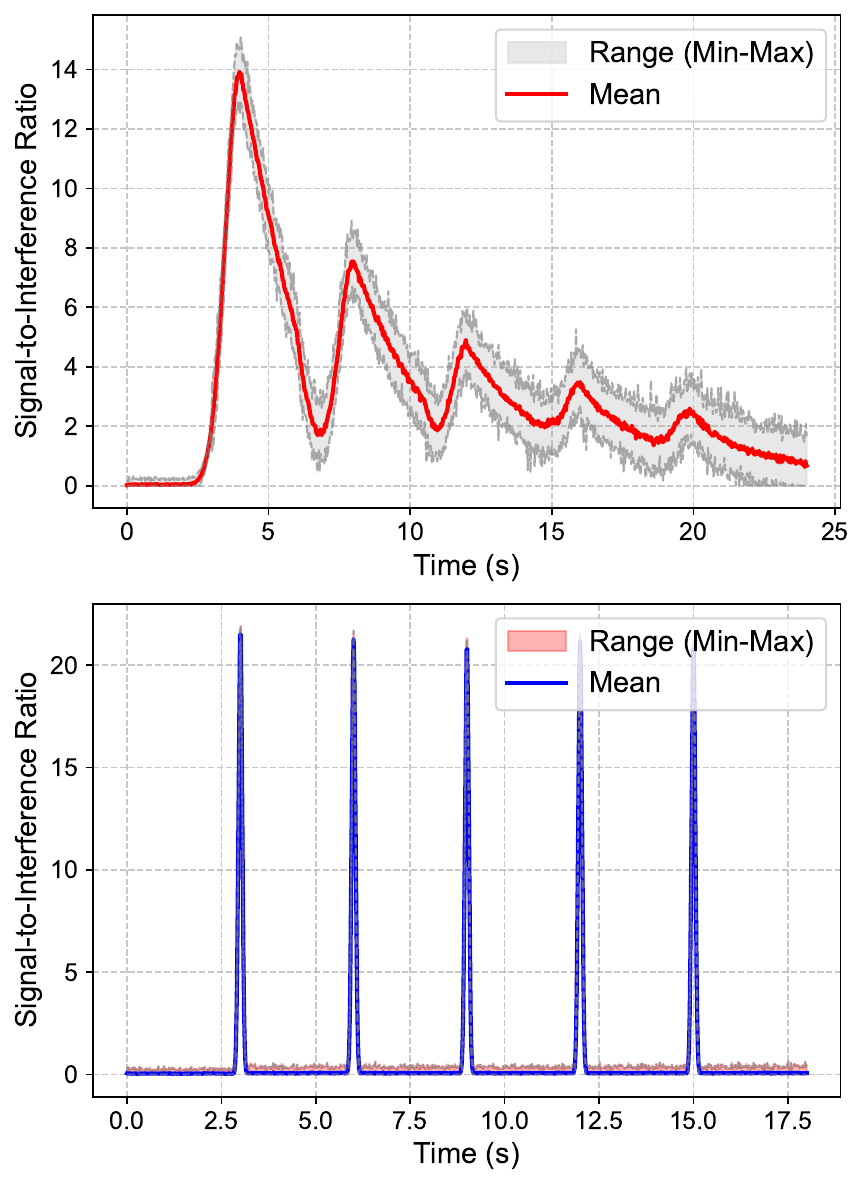}
    \vspace{-10pt}
    \caption{Temporal variations of SIR during the transmission of a continuous sequence of five ‘1’ symbol bits in two molecular communication scenarios.}
    \label{fig:SIR}
    \vspace{-8pt}
\end{figure}

\begin{table}[t]
\caption{Parameters of the Molecular Propagation Channel}
\label{tab:channel-parameters}
\centering
\renewcommand{\arraystretch}{1.2}
\begin{tabular}{|l|c|c|}
\hline
\textbf{Parameter} & \textbf{Scenario 1} & \textbf{Scenario 2} \\ \hline \hline

\textbf{Propagation distance} (\(R\)) 
  & \(100 \,\mathrm{\mu m}\) 
  & \(60 \,\mathrm{cm}\) \\
\hline

\textbf{Receiver radius} (\(r\))  
  & \(20 \,\mathrm{\mu m}\)
  & \(20 \,\mathrm{\mu m}\) \\
\hline

\textbf{Flow velocity} (\(v\))
  & \(50 \,\mathrm{\mu m/s}\)
  & \(40 \,\mathrm{cm/s}\) \\
\hline

\textbf{Symbol duration} (\(t_s\)) 
  & \(4 \,\mathrm{s}\) 
  & \(3 \,\mathrm{s}\) \\
\hline

\textbf{Diffusion coefficient} (\(D_c\)) 
  & \(800 \,\mathrm{\mu m^2/s}\) 
  & \(800 \,\mathrm{\mu m^2/s}\) \\
\hline

\textbf{Maximum released molecules} (\(n_m\)) 
  & \(2 \times 10^4\) 
  & \(2 \times 10^4\) \\
\hline

\end{tabular}
\vspace{-10pt}
\end{table}

\begin{figure}[t]
    \centering
    \includegraphics[width=0.88\linewidth]{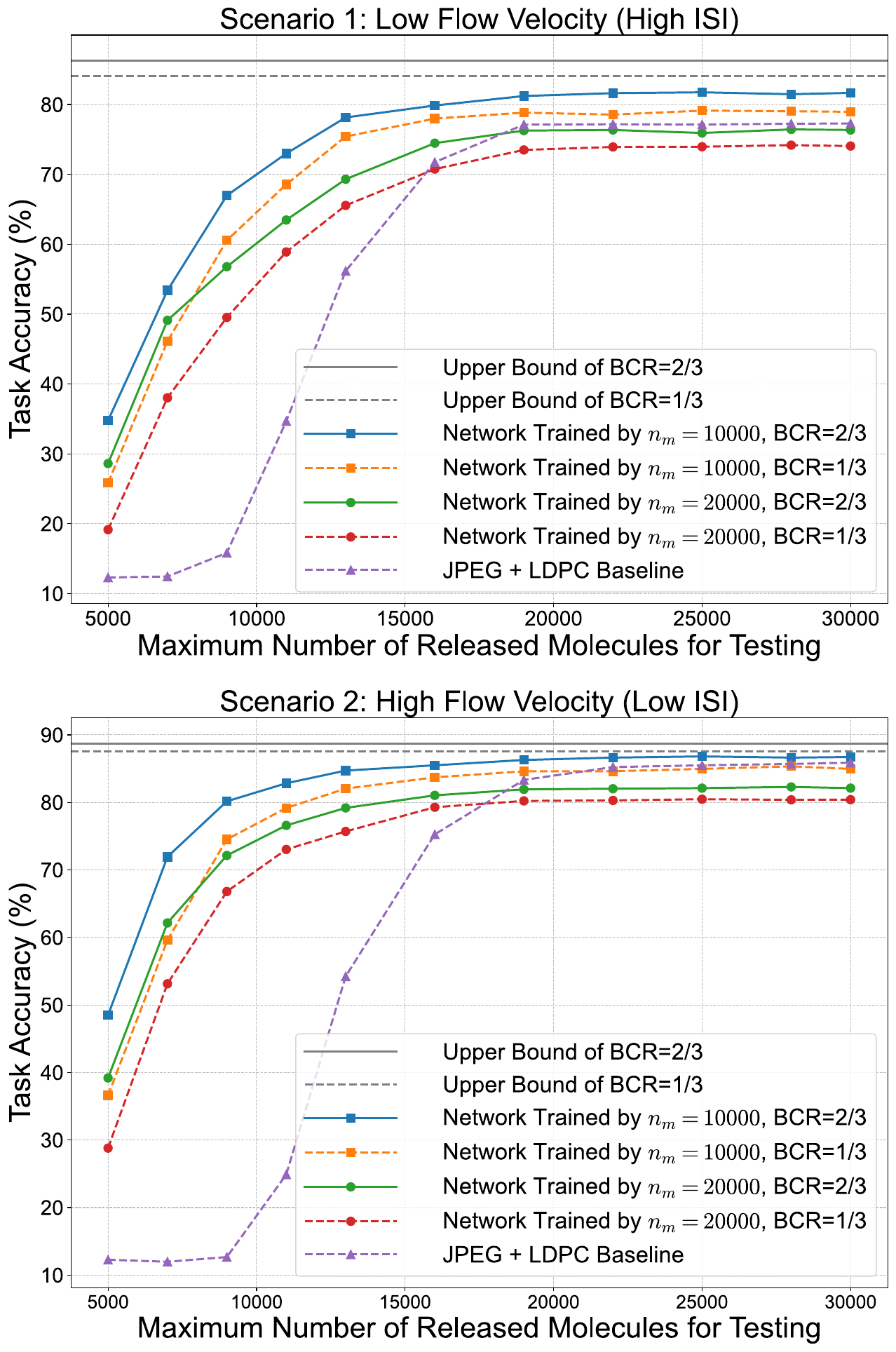}
    \vspace{-6pt}
    \caption{Accuracy performance comparison between different methods in two molecular communication scenarios (BCR: Bandwidth Compression Ratio).}
    \label{fig:accuracy_comparison}
    \vspace{-12pt}
\end{figure}

\section{Experiment and Evaluation}
The proposed framework is evaluated in the context of molecular communication for biomedical image diagnostics using the Kvasir dataset \cite{smedsrud2021kvasir}, which comprises 8,000 high-resolution gastrointestinal images spanning eight clinically significant categories, including normal findings and pathological conditions such as polyps, ulcers, or bleeding. Each image is resized to \( 128 \times 128 \) pixels while preserving RGB channels to retain critical diagnostic features. As shown in Table \ref{tab:channel-parameters}, to simulate practical IoBNT applications, two distinct communication scenarios with different propagation distance, flow velocity and symbol duration are considered \cite{jamali2019channel}.

Fig. \ref{fig:SIR} depicts the temporal variations of SIR during the propagation of five consecutive ‘1’ symbol bits via the trained channel network in two communication scenarios. The results indicate that the proposed channel network closely replicates the physical characteristics of molecular propagation by accurately capturing the pronounced ISI at low flow velocity and the accelerated molecular clearance at high flow velocity. Fig.~\ref{fig:accuracy_comparison} compares the performance of the proposed semantic framework for diagnostic classification tasks under different parameter settings with the conventional JPEG and LDPC-based approach~\cite{yang2022semantic}. The Bandwidth Compression Ratio (BCR) represents the ratio of the compressed feature size to the original input data size, indicating the level of data reduction achieved before transmission. The results show that the proposed method significantly improves accuracy of classification tasks, achieving at least a 25\% performance gain over traditional methods in resource constraint condition (\( n_m < 12,000 \)). Notably, networks trained with lower \( n_m \) values exhibit better learning efficiency, as the increased impact of ISI and channel degradation enables the model to capture more robust propagation features, effectively mitigating the cliff effect observed at low released molecule levels.

\section{Conclusion and Future Work}
This paper proposed an end-to-end semantic molecular communication framework tailored for IoBNT, addressing the challenges of noise, diffusion, and ISI in molecular propagation channels. By introducing a probabilistic channel network, the framework enables joint optimization of the encoder and decoder, ensuring seamless adaptation to stochastic channel conditions. Extensive experiments on diagnostic classification tasks demonstrated that the proposed framework significantly outperforms traditional methods in both accuracy and robustness. Future research will extend the framework to MIMO scenarios, explore adaptive semantic extraction techniques, and enhance efficiency for real-world IoBNT deployments.

% or


% \appendices

% \section{}

% \section*{Acknowledgment}

\ifCLASSOPTIONcaptionsoff
  \newpage
\fi



%\bibliographystyle{IEEEtran}

%%%%%%%%%%%%%%%%%%%%%%%%%%%%%%%%%%%%%%%%%%%%%%%%%%%%%%%%%%%%%%%
%%%%%%%%%%%%%%%%%%%%%%%%%%%%%%%%%%%%%%%%%%%%%%%%%%%%%%%%%%%%%%%

% \newpage
% \newpage

\bibliographystyle{IEEEtran}
% \bibliography{ref}


  % 直接包含 bbl 文件，避免 bibtex 运行

\end{document}